\documentclass{article}
\usepackage{spconf,amsmath,graphicx}
\usepackage{threeparttable}
\usepackage{multirow}
\usepackage{makecell}
\usepackage{url}
\usepackage{cite}
\usepackage{mathrsfs}
\usepackage{multirow}
\usepackage{booktabs}
\usepackage{stfloats}
\usepackage{enumitem}
\usepackage[thicklines]{cancel}
\usepackage{amssymb}
\usepackage{mathrsfs}

\usepackage{amsmath}

\usepackage{CJKutf8}

\title{LAE-ST-MoE: Boosted Language-Aware Encoder Using Speech Translation Auxiliary Task for E2E Code-switching ASR}
\name{Guodong Ma$^{\dag}$, Wenxuan Wang$^{\dag}$, 
Yuke Li$^{\dag*}$\thanks{$^*$ Corresponding author}, 
Yuting Yang$^{\dag}$, Binbin Du$^{\dag}$, Haoran	Fu$^{\ddag}$}
\address{$^{\dag}$NetEase Yidun AI Lab, Hangzhou, China \\
         $^{\ddag}$Department of Civil Engineering, Zhejiang University}
\begin{document}
\copyrightnotice{979-8-3503-0689-7/23/\$31.00~\copyright2023 IEEE}
%

\maketitle
\begin{abstract}
Recently, to 
mitigate
the confusion between different languages in code-switching (CS) automatic speech recognition (ASR), 
the conditionally factorized models, such as the language-aware encoder (LAE), explicitly disregard the contextual information between different languages.
However, this information may be helpful for ASR modeling. 
To alleviate this issue, 
we propose the LAE-ST-MoE framework. It incorporates speech translation (ST) tasks into LAE and utilizes ST to learn the 
contextual information between different languages.
It introduces a task-based mixture of expert 
modules, employing separate feed-forward networks 
for the ASR and ST tasks.
Experimental results on the ASRU 2019 Mandarin-English CS challenge dataset demonstrate that, compared to the LAE-based CTC, the LAE-ST-MoE model achieves a 9.26\% mix error reduction on the CS test with the same 
decoding parameter.
Moreover, the well-trained LAE-ST-MoE model can perform ST tasks from CS speech to Mandarin or English text.

\end{abstract}
\section{Introduction}
\label{sec:intro}
With the rise of end-to-end (E2E) automatic speech recognition (ASR), researchers \cite{ctc-paper,RNNt-paper,LAS-paper,CTC_conformer,Speech-Transformer,Conformer,SpecAugment,yuting_paper,ma22_interspeech,policAugment_ma,code-switch_paper10,code-switch_paper11,code-switch_paper2,code-switch_paper1,code-switch_paper7,code-switch-non-auto,code-switch_paper4,code-switch_paper5,code-switch_paper8,code-switch-bytedance,2020Bi,wang23_interspeech,you2021speechmoe,mole_paper,ma21_interspeech}  explore different E2E ASR scenarios.
An utterance that includes two or more languages is known as a code-switching (CS) scenario,
which is generally divided into occurring at an utterance level (extra-sentential CS) or within an utterance (intra-sentential CS). It is still a challenging ASR scenario.

Several challenges are conventionally encountered in modeling CS speech: firstly, 
the real paired CS audio is data-scarce,
and secondly, 
the conventional models are not good at modeling CS speech due to the confusion between different languages.
To alleviate the first issue, researchers propose technical methods to study the rules of CS occurrence and synthesize CS paired data \cite{code-switch_paper2,code-switch_paper1,code-switch_paper7,code-switch_paper10,code-switch_paper11} or explore the affection of monolingual data \cite{code-switch_paper6,code-switch_paper9,code-switch-non-e2e-pair}.
As for the second issue, the structures like Connectionist Temporal Classification (CTC)-, attention-, and transducer-based E2E models have been investigated for CS ASR \cite{code-switch_paper1,code-switch_paper2,code-switch-non-auto,code-switch_paper4,code-switch_paper5,code-switch_paper7,code-switch_paper8,2020Bi,wang23_interspeech,code-switch-bytedance}.
Recently, to mitigate the second issue, the conditionally factorized frameworks \cite{zero-shot_cs,yan2022joint,Song2022LanguagespecificCA,tian22c_interspeech} are proposed to decompose the CS task (e.g., Mandarin-English CS) into two modeling steps: 1) recognizing Mandarin 
and English part, respectively, and 2) composing processed monolingual segments into a CS sequence.
However, in modeling step 1) for these methods, the model only utilizes the information of the monolingual part. We know that, when modeling the non-streaming E2E ASR task, the prediction of each unit generally relies on overall audio contextual information. 

To solve the issues of the conditionally factorized models 
\cite{zero-shot_cs,yan2022joint,Song2022LanguagespecificCA,tian22c_interspeech}
(e.g., LAE \cite{tian22c_interspeech}),
we propose the LAE-ST-MoE framework. It incorporates speech translation (ST) tasks into LAE \cite{tian22c_interspeech} and utilizes ST to facilitate the learning of contextual information between Mandarin and English, thereby impacting the model’s encoder through joint learning. 
In addition, inspired by \cite{you2021speechmoe,wang23_interspeech,mole_paper}, the LAE-ST-MoE 
introduces a task-based mixture of expert (MoE) approach, employing separate feed-forward networks (FFNs) for the ASR and ST tasks. 

Our experiment is conducted on the classic CS benchmark, i.e., ASRU 2019 Mandarin-English CS challenge dataset \cite{shi2020asru}. Since the data does not have ST labels, 
we use the large machine translation (MT) model from ModeScope 
to label the data, which is based on the CSANMT algorithm \cite{model_scope_nmt}.
In the experiments, compared to the LAE-based system, the LAE-ST-MoE model achieves a relative performance improvement of about 6\%-9\% in ASR tasks on all test sets. Moreover, our model does not introduce extra decoding computational complexity.
In addition, 
the trained LAE-ST-MoE model can perform ST tasks from CS speech to Mandarin or English text and has achieved good BLEU. Then, it is easy to extend our model to one-to-many ST tasks.

Our main contributions are as follows:
(1) To our best knowledge, we are the first to propose using the ST task to introduce richer cross-lingual contextual information to boost the monolingual modeling stage of LAE;
(2) We introduce 
an MoE between ASR and ST tasks to make each task more focused, thereby improving the overall recognition performance of the model without extra decoding computational complexity; 
(3) The well-trained LAE-ST-MoE model can perform ST tasks from CS speech to Mandarin or English text, and the structure is easy to extend to one-to-many ST tasks. 
\vspace{-0.3 em}
\section{Problem formulations and Motivation}
\vspace{-0.3 em}
\label{sec:relate_work_movitvation}


\label{ssec:lae_framework}

In the Mandarin-English CS ASR system \cite{zero-shot_cs,yan2022joint,Song2022LanguagespecificCA,tian22c_interspeech}, we know that the basis is to model the label-to-frame alignments. For each T-length speech feature sequence $\rm{X} = \{\rm{x_t} | t = 1, ..., T\}$ and L-length CS label sequence $\rm{Y} = \{y_{\ell} \in (\rm \mathbb{V}^{Man} \bigcup \mathbb{V}^{En} | \ell = 1, ..., L)\}$, there are several possible T-length label-to-frame sequences $\rm{Z} = \{ z_t \in (\mathbb{V}^{Man} \bigcup \mathbb{V}^{En} \\ \bigcup \{ \emptyset \} ) | t = 1, ..., T \}$, where $\emptyset$ denotes a blank symbol in CTC \cite{ctc-paper} based
system, $\rm \mathbb{V}^{Man}$, and $\rm \mathbb{V}^{En}$ respectively represents to the Mandarin and English part in CS. However, for each
CS Z, there always are two corresponding monolingual label-to-frame sequences $\rm Z^{Man} = \{ z_t^{Man} \in \{\mathbb{Z}^{Man} \bigcup \{\emptyset\}\} | t = 1, ..., T\}$ and $\rm Z^{En} = \{ z_t^{En} \in \mathbb{Z}^{En} \bigcup \{\emptyset\}\} | t = 1, ..., T\}$. Therefore, the label-to-frame posterior $\rm P(Y | X)$ can thus be represented in terms of CS, $\rm P(Z | X)$, and monolingual, $\rm P(Z^{Man} | X)$ and $\rm P(Z^{En} | X)$, label-to-frame posteriors: 
\begin{eqnarray}
\label{eq:bilingual_eq}
    \rm P ({\rm Y}| {\rm X)} = \sum_{Z \in \mathbb{Z}} \sum_{Z^{Man} \in \mathbb{Z}^{Man}} \sum_{Z^{En} \in \mathbb{Z}^{En}} P(Z, Z^{Man}, Z^{En}| \rm X)
\end{eqnarray}
where $\mathbb{Z}$ and $\rm \mathbb{Z}^{Man/En}$ denote sets of all possible CS and monolingual label-to-frame alignments for a given Y. By applying Bayes' formula, the $\rm P(Z, Z^{Man}, Z^{En}| \rm X)$ in Eq.(\ref{eq:bilingual_eq}) can be transformed into the following expression: 
\vspace{-1.0 em}
\begin{eqnarray}
\label{eq:bilingual_eq_viriet1}
\begin{aligned}
    \rm P(Z, Z^{Man}, Z^{En}| \rm X) = P(Z| Z^{Man}, Z^{En}, \rm X) \\ \times \rm P(Z^{Man}, Z^{En}| \rm X) 
\end{aligned}
\end{eqnarray}
\vspace{-1.0 em}
and 
\begin{eqnarray}
\label{eq:bilingual_eq_viriet2}
    \rm P(Z^{Man}, Z^{En}| \rm X) = P( Z^{Man} | Z^{En}, \rm X) \times P(Z^{En}| \rm X).
\end{eqnarray}

\begin{figure*}[t]
\centering
 \centerline{\includegraphics[width=16cm]{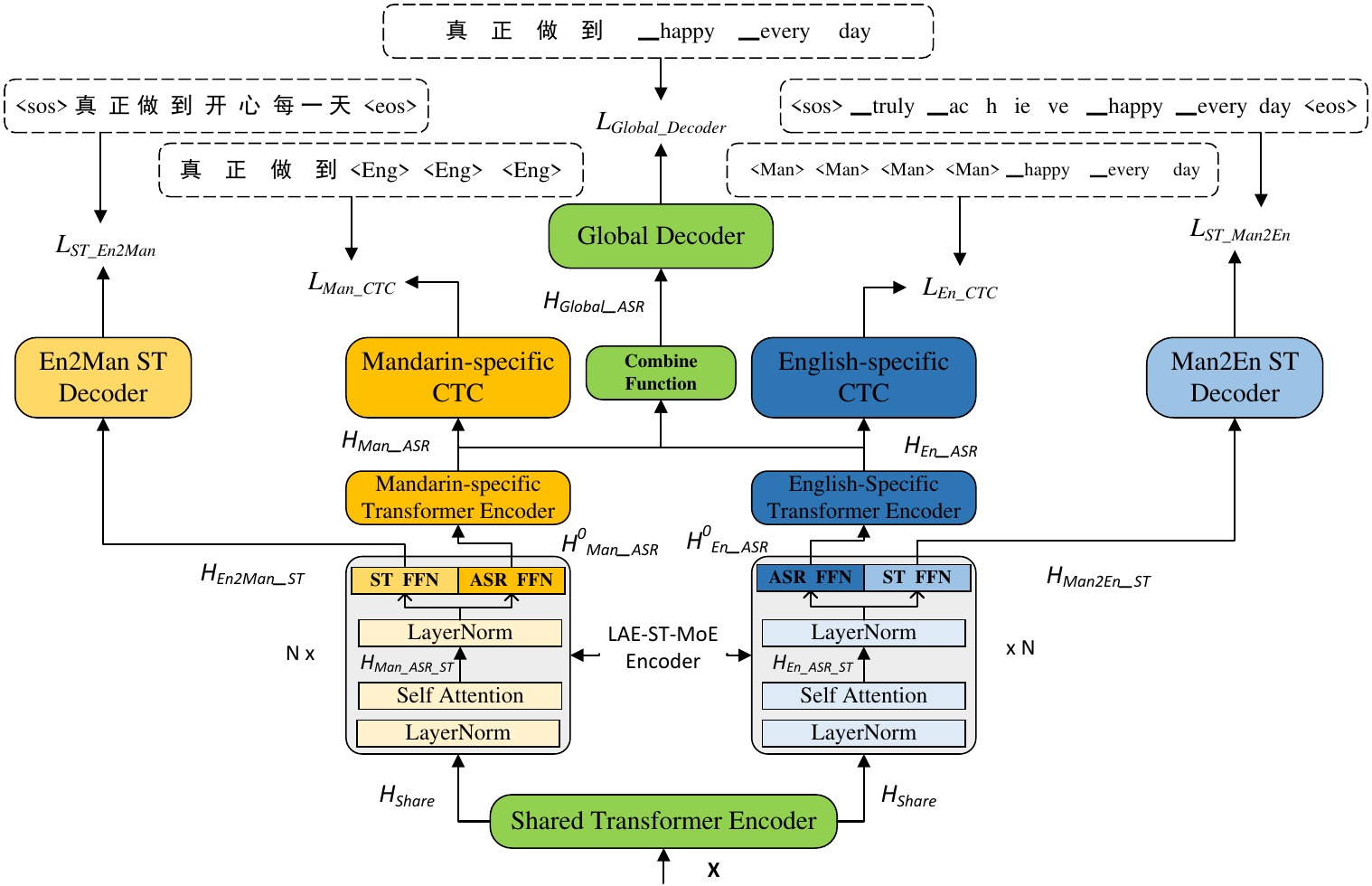}}
\caption{The framework of the proposed LAE-ST-MoE.}
\label{fig:lae_st_moe}
 
\end{figure*}

Two assumptions are made. The first assumption is that once $\rm Z^{Man}$ and $\rm Z^{En}$ are given, no additional information from observation X is needed to determine Z.
The second assumption is that $\rm Z^{Man}$ and $\rm Z^{En}$ are independent, given X.
Therefore, combined with Eq.~(1-3), the eq.~(\ref{eq:bilingual_eq}) can be shown:
\vspace{-0.5 em}
\begin{eqnarray}
\setlength\tabcolsep{1.0pt}
\label{eq:conditional_eq}
  \begin{aligned}
    \rm P ({\rm Y}| {\rm X)} \approx \sum_{Z \in \mathbb{Z}} P(Z | Z^{Man}, Z^{En}) \times \sum_{Z^{Man} \in \mathbb{Z}^{Man}} \hspace{-0.2cm} P( Z^{Man} | \rm X) \\
    \rm \times \sum_{Z^{En} ~\in~\mathbb{Z}^{En}~~~} \hspace{-0.2cm} P( Z^{En}|\rm X).
  \end{aligned}
\end{eqnarray}
To achieve the transformation from Eq. (\ref{eq:bilingual_eq}) to Eq. (\ref{eq:conditional_eq}), the monolingual-specific encoder is introduced by the conditionally factorized structures \cite{zero-shot_cs,yan2022joint,Song2022LanguagespecificCA,tian22c_interspeech} to optimize the representation of each language separately. For example, the token sequence of the 
 CS audio is like “\begin{CJK*}{UTF8}{gbsn}
真~~正~~做~~到
\end{CJK*}~$\rm \_happy$~~$\rm \_every$~~\\$\rm day$". When forwarding Mandarin-specific encoder, the reference text will be replaced with “\begin{CJK*}{UTF8}{gbsn}
真~~正~~做~~到
\end{CJK*}~~$<$$\rm En\_tok$$>$~~\\$<$$\rm En\_tok$$>$ $<$$\rm En\_tok$$>$" and ignore the English part, where 
 $<$$\rm En\_tok$$>$ can refer to $<$UNK$>$ \cite{Song2022LanguagespecificCA} or $<$Eng$>$ \cite{tian22c_interspeech}.
As for the English-specific encoder, as shown in Figure~\ref{fig:lae_st_moe}, it is the same as the Mandarin-specific encoder. Further consider the modeling process, e.g., Mandarin-specific encoder, 
the model will not learn English contextual information in the CS audio, which could potentially improve its performance on the Mandarin part.
However, 
the ST model is capable of converting contextual information from various languages into one language.
Therefore, applying ST tasks to enrich the contextual information between the two languages in CS ASR can be reasonable and feasible. Based on the LAE architecture \cite{tian22c_interspeech} and joint learning mechanism, we propose LAE-ST-MoE architecture, which uses ST as an auxiliary task to bring more contextual information for ASR. 
The details of our proposed LAE-ST-MoE will be presented in the next section. 

\vspace{-0.5 em}

\vspace{-0.3 em}
\section{Proposed Frameworks}
\vspace{-0.3 em}
\label{sec:pagestyle}

\subsection{LAE-ST-MoE architecture}
\label{ssec:lae_st_moe}
The LAE structure \cite{tian22c_interspeech} has a shared encoder module, two language-specific encoders for Mandarin and English, and a global ASR decoder. The monolingual-specific encoder is imposed by a corresponding monolingual-specific CTC loss. To alleviate the issues of LAE discussed in section~\ref{ssec:lae_framework}, we propose the LAE-ST-MoE model architecture, as shown in Figure~\ref{fig:lae_st_moe}, which introduces two LAE-ST-MoE encoders and two ST decoders based on LAE.
If $\mathbf{N_{Share}}$ represents the number of layers in the shared encoder, and $\mathbf{N_{Mono}}$ represents the number of layers in the monolingual-specific encoder.
Then, the LAE-ST-MoE encoder has $\mathbf {N}$ layers, where $\mathbf {N}$ is equal to ($\mathbf {N_{Encoder}} - \mathbf{N_{Share}} - \mathbf {N_{Mono}}$) and $\mathbf {N_{Encoder}}$ refers to the overall encoder layers. 
A common ST cross-entropy loss imposes the ST decoder, which consists of 6 Transformer-based blocks.
In addition, the ST and ASR tasks are jointly trained using FFN-based MoE. A detailed explanation of the proposed LAE-ST-MoE model is presented as follows. 

If given the input feature sequence X, the shared Transformer encoder will transform it to representation $\mathbf{H_{share}}$:
\begin{eqnarray}
\label{eq:share_en_out}
    \mathbf{H_{share}} = \rm Encoder_{share} (\rm X).
\end{eqnarray}
Furthermore, the $\mathbf{H_{share}}$ will be forwarded to the LAE-ST-MoE encoder, which replaces the FFN of the Transformer encoder with the FFN-MoE module. It produces the hybrid ASR-ST representation $\mathbf{H_{Man\_ASR\_ST}}$ and $\mathbf{H_{En\_ASR\_ST}}$ using multi-head self-attention (MHSA):
\begin{eqnarray}
\label{eq:asr-st_en_out}
    \mathbf{H_{Man\_ASR\_ST}} = \rm  MHSA(LNorm(\mathbf{H_{share}})) \\
    \mathbf{H_{En\_ASR\_ST}} = \rm MHSA(LNorm(\mathbf{H_{share}}))
\end{eqnarray}
where LNorm denotes LayerNorm \cite{layer_norm_paper}. 
Based on $\mathbf{H_{En\_ASR\_ST}}$ and $\mathbf{H_{Man\_ASR\_ST}}$, the FFN-MoE is forward to get ASR representation $\mathbf{H^{0}_{En\_ASR}}$, $\mathbf{H^{0}_{Man\_ASR}}$, and ST representation $\mathbf{H_{Man2En\_ST}}$, $\mathbf{H_{En2Man\_ST}}$, respectively: 
\begin{eqnarray}
\label{eq:fnn-moe_out}
    \mathbf{H^{0}_{Man\_ASR}} = \rm LNorm(FFN\_MoE(\mathbf{H_{Man\_ASR\_ST}})) \\
    \mathbf{H_{En2Man\_ST}} = \rm LNorm(FFN\_MoE(\mathbf{H_{Man\_ASR\_ST}})) \\
    \mathbf{H^{0}_{En\_ASR}} = \rm LNorm(FFN\_MoE(\mathbf{H_{En\_ASR\_ST}})) ~~~\\
    \mathbf{H_{Man2En\_ST}} = \rm LNorm(FFN\_MoE(\mathbf{H_{En\_ASR\_ST}})). ~~
\end{eqnarray}

On the ST task side, $\mathbf{H_{En2Man\_ST}}$ and $\mathbf{H_{Man2En\_ST}}$ will forward to the En2Man and Man2En ST decoder, respectively. In ASR, it is the same as LAE, based on $\mathbf{H^{0}_{Man\_ASR}}$ and $\mathbf{H^{0}_{En\_ASR}}$, the Monolingual-specific encoder will produce the monolingual-specific representation $\mathbf{H_{Man\_ASR}}$, $\mathbf{H_{En\_ASR}}$ and combine these to get the global ASR representation $\mathbf{H_{Global\_ASR}}$:
\begin{eqnarray}
\label{eq:encoder_asr_out}
    \mathbf{H_{Man\_ASR}} &=& \rm Encoder_{Man\_Spec}(\mathbf{H^{0}_{Man\_ASR}}) \\
    \mathbf{H_{En\_ASR}} &=& \rm Encoder_{En\_Spec}(\mathbf{H^{0}_{En\_ASR}}) \\
    \mathbf{H_{Global\_ASR}} &=& \rm \mathbf{H_{Man\_ASR}} + \mathbf{H_{En\_ASR}} .
\end{eqnarray}

\subsection{Training and Decoding}
\label{ssec:lae_st_moe_train_decode}
In the  LAE-ST-MoE model training stage, if the label text sequence for speech feature X is $\rm Y$, we will apply the $\rm Model_{MT}$ from ModelScope to translate $\rm Y$ into Mandarin $\rm Y^{Man}$ and English $\rm Y^{En}$ text:
\begin{eqnarray}
\label{eq:mt_out}
    \rm Y^{Man} &=& \rm Model_{MT\_En2Man}(Y) \\
    \rm Y^{En} &=& \rm Model_{MT\_Man2En}(Y).
\end{eqnarray}

Like \cite{tian22c_interspeech,Song2022LanguagespecificCA}, we replace $\rm Y$ with monolingual-specific label $\rm Y^{Man\_Spec}$ and $\rm Y^{En\_Spec}$ using $<$Eng$>$ and $<$Man$>$, respectively. Based on monolingual-specific ASR representation $\mathbf{H_{Man\_ASR}}$ and $\mathbf{H_{En\_ASR}}$, the monolingual-specific ASR object $\mathcal{L}_{Spec}$ will be shown as follow:
\begin{eqnarray}
\label{eq:spec_loss_out}
    \rm \mathcal{L}_{Man\_CTC} = \rm CTC_{Man\_Spec} (Y^{Man\_Spec} | \mathbf{H_{Man\_ASR}}) \\
    \rm \mathcal{L}_{En\_CTC} = \rm CTC_{En\_Spec} (Y^{En\_Spec} | \mathbf{H_{En\_ASR}}) \quad~~~\\
    \rm \mathcal{L}_{Spec} = \frac{( \mathcal{L}_{Man\_CTC} + \mathcal{L}_{En\_CTC} )}{2}. \quad\quad\quad\quad~~~~~
\end{eqnarray}
Moreover, given the global ASR representation $\rm \mathbf{H_{Global\_ASR}}$, the global ASR decoder object $\rm \mathcal{L}_{Global\_Decoder}$ is:
\begin{eqnarray}
\label{eq:global_loss_out}
    \rm \mathcal{L}_{Global\_Decoder} = \rm Decoder_{Global} (Y | \mathbf{H_{Global\_ASR}}).
\end{eqnarray}

Following \cite{Song2022LanguagespecificCA}, we also use $\rm \lambda_{Spec}$ (we set it to 0.3 in the experiments) to combine $\rm \mathcal{L}_{Spec}$ and $\rm \mathcal{L}_{Global\_Decoder}$ to produce the overall ASR loss $\mathcal{L}_{\rm ASR}$:
\begin{eqnarray}
\label{eq:overall_asr_loss_out}
   \rm \mathcal{L}_{ASR} \hspace{-0.1cm} = \hspace{-0.1cm} \rm \lambda_{Spec} \hspace{-0.1cm} \times \mathcal{L}_{Spec}  + (1 - \lambda_{Spec}) \times \mathcal{L}_{Global\_Decoder}.
\end{eqnarray}
In the CTC-based ASR system, $\rm \mathcal{L}_{Global\_Decoder}$ only represents the CTC loss. Otherwise, in hybrid CTC attention-based ASR \cite{CTC_conformer}, $\rm \mathcal{L}_{Global\_Decoder}$ is the combination between CTC $\rm \mathcal{L}_{Global\_CTC}$ and attention $\rm \mathcal{L}_{Global\_Att}$ loss using $\rm \lambda_{CTC}$: 
\begin{eqnarray}
\begin{aligned}
\label{eq:hybrid_ctc_loss_out}
   \rm \mathcal{L}_{Global\_Decoder} &=& \rm \mathcal{L}_{Global\_CTC} \times \lambda_{CTC} \quad\quad\quad \\ \rm &&+ (1  - \lambda_{CTC}) \times \mathcal{L}_{Global\_Att}.
\end{aligned}
\end{eqnarray}

On the ST task, given ST representation ($\mathbf{H_{En2Man\_ST}}$ and $\mathbf{H_{Man2En\_ST}}$) and ST label sequence ($\rm Y^{Man}$ and $\rm Y^{En}$),  the overall ST loss $\rm \mathcal{L}_{ST}$ is shown as follows: 
\begin{eqnarray}
\label{eq:st_loss_out}
    \rm \mathcal{L}_{ST\_Man2En} = \rm Decoder_{Man2En} (Y^{En} | \mathbf{H_{Man2En\_ST}}) \\
    \rm \mathcal{L}_{ST\_En2Man} = \rm Decoder_{En2Man} (Y^{Man} | \mathbf{H_{En2Man\_ST}}) \\
    \rm \mathcal{L}_{ST} = \frac{( \mathcal{L}_{ST\_Man2En} + \mathcal{L}_{ST\_En2Man} )}{2} \quad\quad
\end{eqnarray}
where we use the cross-entropy loss for the ST tasks.

Based on the overall ASR loss $\rm \mathcal{L}_{ASR}$ and ST loss $\rm \mathcal{L}_{ST}$, the final training object $\rm \mathcal{L}_{Final}$ is:
\begin{eqnarray}
\label{eq:final_loss_out}
    \rm \mathcal{L}_{Final} = \mathcal{L}_{ASR} + \beta \times \mathcal{L}_{ST}
\end{eqnarray}
where $\beta$ is used to balance and regulate the ST effect.

In the ASR decoding stage, like the LAE structure \cite{tian22c_interspeech}, our model only gets the probabilities from the global ASR decoder. Therefore, compared with \cite{tian22c_interspeech}, our LAE-ST-MoE model has the same decoding computational complexity. In the ST decoding, our model uses the custom auto-regressive manner to forward the corresponding ST branch and get the final ST results. 
In addition, for monolingual Mandarin input, the En2Man ST decoder is comparable to the Mandarin ASR decoder. Therefore, we can easily fuse it into monolingual Mandarin decoding through rescoring. The same applies to monolingual English decoding. 

\vspace{-1.0 em}

\begin{table}[ht] 
\setlength\tabcolsep{4.0pt}
  \caption{The details of the used Datasets}
  \vspace{0.5 em}
  \label{tab:asru_data}
    \center
    \begin{tabular}{c c | c c | c c} \hline
    \multicolumn{1}{c}{\multirow{2}{*}{Lang}} & \multicolumn{1}{c|}{\multirow{2}{*}{Corpora}} & \multicolumn{2}{c|}{Dur. (Hrs)} & \multicolumn{2}{c}{Utterance(k)} \\ \cline{3-6}
      &   & Train & Eval & Train & Eval \\ 
    \hline
    CN & ASRU-Man\cite{shi2020asru} & 482.6 & 14.3 & 545.2 & 16.6\\ 
    EN & Librispeech\cite{panayotov2015librispeech} & 464.2 & 10.5 & 132.5 & 5.6 \\
    CN-EN & ASRU-CS\cite{shi2020asru} & 199.0 & 20.3 & 186.4 & 16.2\\
    \hline
    \end{tabular}
    \vspace{-2.0 em}
\end{table}

\section{Experiments and results}
\label{sec:typestyle}

\subsection{Datasets}
\label{ssec:dataset}

We experiment on ASRU 2019 Mandarin-English code-switching challenge dataset\cite{shi2020asru}. Like \cite{wang23_interspeech}, we split the same Mandarin monolingual subset of the ASRU 2019 dataset as our CN test. 
Moreover, we use the test-clean and test-other datasets from 
Librispeech \cite{panayotov2015librispeech} to create our monolingual English test EN.
Then, 
the CS test CN-EN is from the official 
challenge test set.
The details are presented in Table~\ref{tab:asru_data}.



The 80-dimensional log filter-bank energy is our input acoustic features, which are extracted with a stride size 10ms and a window size 25ms. The cepstral mean and variance normalization (CMVN), and 
SpecAugment \cite{park2019specaugment} is applied. 
The 
vocabulary consists of 7075 unique characters and 4989 BPE \cite{sennrich2015neural} tokens. In addition, as for the training and testing ST label, the EN2CN\footnote{https://www.modelscope.cn/models/damo/nlp\_csanmt\_translation\_en2\\zh/summary} and CN2EN\footnote{https://www.modelscope.cn/models/damo/nlp\_csanmt\_translation\_zh2\\en/summary} translation model, which is based on the CSANMT algorithm \cite{model_scope_nmt}, both from ModelScope\footnote{https://github.com/modelscope/modelscope}, is used to get the pseudo labels. Then, we use WeNet's \cite{wenet_paper} metrics calculation script\footnote{https://github.com/wenet-e2e/wenet/blob/main/tools/compute-wer.py} for ASR scoring, which includes word (WER), character (CER), mix (MER) error rate, and the sacrebleu \cite{sacrebleu_paper} tool for ST scoring, which includes BLEU and translation error rate (TER). 

For simpler expression, in Table~\ref{tab:main_res}, Table~\ref{tab:ab_w_w/o_moe}, Table~\ref{tab:res_st_rescore}, Table~\ref{tab:res_asr_beta}, and Table~\ref{tab:res_asr_N_mono}, we will use CN, EN, and ALL to represent the CER of monolingual Mandarin, the WER of monolingual English, and the total MER of the CS test set respectively.

\vspace{-0.5 em}
\subsection{Experimental setup}
\label{ssec:exp_setup}

The experiments are both conducted on the ESPnet toolkit \cite{watanabe2018espnet}. We use the 
 hybrid CTC/Attention \cite{CTC_conformer} model with a $\mathbf{N_{Encoder}}$=12 encoder, $\mathbf {N_{Decoder}}$=6 decoder, and the CTC-only model with a $\mathbf{N_{Encoder}}$=12, called the Vallina model. In the hybrid CTC/Attention model, $\rm \lambda_{CTC}$ set to 0.3. 
In our implementation, following \cite{tian22c_interspeech}, the LAE-based baseline model contains a shared encoder block $\mathbf {N_{Share}}$=9 and a language-specific encoder block $\mathbf {N_{Mono}}$=3 for each language. As mentioned in section~\ref{ssec:lae_st_moe}, the layers of the LAE-ST-MoE encoder $\mathbf{N}$ are equal to ( $\mathbf{N_{Encoder}}$ - $\mathbf {N_{Share}}$ - $\mathbf {N_{Mono}}$ ), and the number of layers will be given in the result section. 
In our models, all encoders and decoders are stacked Transformer-based blocks \cite{attention_is_all_you_need,Speech-Transformer} with an attention dimension of 256, 4 attention heads, and a feed-forward dimension of 2048. 

We use the Adam optimizer with a Transformer-lr scale of 1 and warmup steps of 25k to train 100 epochs on 8 Tesla V100 GPUs. 
The dropout rate is 0.1 
to prevent the model from over-fitting. 
In the training stage, we adopt a dynamic batch size strategy with a maximum batch size of 128. Moreover, we use Kenlm \cite{kenlm_paper} to train a 4-gram language model with all training transcriptions and adopt the CTC prefix beam search for ST decoder rescore with a fixed beam size 10. 


\begin{table*}[t]
\renewcommand\tabcolsep{4.0pt}
 \centering
 \caption{Results of proposed models and the baselines. The numbers in brackets indicates the relative error rate reduction comparing with the corresponding LAE-based model (S2 and S5). 
 }
 \vspace{0.5 em}
 \label{tab:main_res}
 \begin{tabular}{l l c | c c  c | c | c} \hline
 \multicolumn{1}{l}{\multirow{2}{*}{\textbf{System}}} &
    \multicolumn{1}{l}{\multirow{2}{*}{\textbf{Model}}} &
    \multicolumn{1}{c|}{\multirow{2}{*}{\textbf{Infer Params}}} &
    \multicolumn{3}{c|}{\textbf{Code-Switch}}&
    \multicolumn{2}{c}{\textbf{Mono}} \\ \cline{4-8}
     & & & \textbf{ALL} & \textbf{CN} & \textbf{EN}  & \textbf{EN} & \textbf{CN} \\
     \hline
     \multicolumn{8}{l}{CTC-based ASR system} \\
     \hline
      Literature &  &  &  &  &  &  &  \\
     - & Conformer CTC \cite{tian22c_interspeech} &  - & 11.6 & - & - & - & -  \\
       & ~ + LAE \cite{tian22c_interspeech} &  - & \textbf{9.5} & - & - & - & -  \\
     - & FLR-MoE CTC \cite{wang23_interspeech} &  25.8 M & 10.5 & 7.7 & 33.1 & 10.1 & 5.1 \\
     \hline
     Our results &  &  &  &  &  &  &  \\
      S1 & Vallina CTC & 19.8 M & 12.2 & 9.0 & 38.9 & 12.4 & 7.1  \\ 
     S2 & LAE CTC (baseline)      & 26.5 M & 10.8 & 8.0 & 33.7 & 10.5 & 5.3  \\
     S3 & LAE-ST-MoE CTC (proposed) & 26.5 M & \textbf{9.8 (9.26\% $\downarrow$)} & \textbf{7.3} & \textbf{30.3 } & \textbf{9.6 (8.57\% $\downarrow$)} & \textbf{4.9 (7.55\% $\downarrow$)}  \\
     \hline
     \multicolumn{8}{l}{Attention-based ASR system } \\
     \hline
     Literature &  &  &  &  &  &  &  \\
     - & Hybrid CTC + Attention \cite{2020Bi} &  28.8 M & 10.9 & 8.8 & 28.1 & - & -  \\
       & ~ + Bi-En. (MoE-in-unsup) \cite{2020Bi} &  45.6 M & 9.8 & 7.7 & 26.6 & - & -  \\
      - & FLR-MoE AED \cite{wang23_interspeech} &  40.7 M & 9.7 & 7.4 & 28.4 & 9.6 & \textbf{4.7} \\
    \hline
     Our results &  &  &  &  &  &  &  \\
     S4 & Vallina AED & 34.7 M & 11.2 & 8.6 & 32.5 & 11.7 & 6.3 \\
     S5 & LAE AED (baseline) & 41.4 M & 10.0  & 7.7 & 29.2 & 9.9 & 5.0 \\
     S6 & LAE-ST-MoE AED (proposed)  & 41.4 M & \textbf{9.3 (7\% $\downarrow$)} & \textbf{7.1} & \textbf{27.4} & \textbf{9.2 (7.07\% $\downarrow$)} &  \textbf{4.7 (6\% $\downarrow$)} \\
     \hline
 \end{tabular}


\end{table*}

\vspace{-1.0 em}

\subsection{Experimental Results}
\label{ssec:results}

\subsubsection{Main results}
\label{sssec:comp_with_lae}
To show the effectiveness of our proposed LAE-ST-MoE framework, we compare it with LAE-based CTC and attention-based (AED) ASR models. We set the $\mathbf {N_{Mono}}$ to 1 and $\rm \beta$ to 0.6 in these experiments. The ablation on $\rm \beta$ and $\mathbf {N_{Mono}}$ will be shown in section~\ref{sssec:ab_w_st_ps}. The results are shown in Table~\ref{tab:main_res}.
\\ \textbf{CTC System:} Compared with the LAE-CTC ASR system (S2), our proposed LAE-ST-MoE CTC model (S3) achieve 9.26\%, 8.57\%, and 7.55\% relative performance gain over the CS, mono EN, and CN tests, respectively, with the same decoding parameter. Especially in the English part of the CS test, our LAE-ST-MoE CTC (S3) shows a 10.09\% WER reduction over the LAE CTC (S2) system. Moreover, it demonstrates a superior performance gain compared to Vanilla CTC (S1), which shows an about 20\% error rate reduction in the CS test. Furthermore, the proposed LAE-ST-MoE CTC achieves a comparable performance with Conformer-based LAE \cite{tian22c_interspeech} and an obvious gain compared to FLR-MoE CTC \cite{wang23_interspeech}.
\\ \textbf{AED System:} The results also show that our LAE-ST-MoE-based system (S6) performs better than the Vallina (S4) and LAE-based (S5) AED ASR. Moreover, the LAE-ST-MoE-based AED system (S6) also shows an obvious MER reduction compared with the Bi-encoder \cite{2020Bi} based and FLR-MoE \cite{wang23_interspeech} based system on the CS test. 
\\ \textbf{CTC vs. AED system:} We can find that the proposed LAE-ST-MoE-based CTC (S3) shows a little performance gain to the LAE AED system (S5) and comparable results with Bi-Encoder \cite{2020Bi} based and FLR-MoE \cite{wang23_interspeech} based AED system. 

These results suggest that the ST auxiliary task can improve the ASR performance based on the LAE structure, which is consistent with our motivation.

\subsubsection{Results of the w/ or w/o MoE in LAE-ST-MoE model}
\label{sssec:ab_w_w/o_moe}

\begin{table}[ht!] 
  \renewcommand\tabcolsep{1.4pt}
  \begin{threeparttable}
  \caption{Performance of the w/ or w/o MoE.
  } 
  \vspace{-1.5 em}
  
  \label{tab:ab_w_w/o_moe}
    \center

    \begin{tabular}{l | c c c | c | c| c | c} \hline
    \multicolumn{1}{c|}{\multirow{2}{*}{\textbf{Model}}} &
    \multicolumn{3}{c|}{\textbf{Code-Switch}} &
    \multicolumn{2}{c|}{\textbf{Mono}} &
    \textbf{CS → EN} & 
    \textbf{CS → CN} \\ \cline{2-8}
     & \textbf{ALL} &  \textbf{CN} & \textbf{EN} & \textbf{EN} & \textbf{CN} & \textbf{BLEU} & \textbf{BLEU} \\ 
     \hline  
    LAE-ST CTC & 10.0 & 7.4 & 31.6 & 9.8 & 5.2 & 16.2 & 65.8\\
    \quad + MoE & \textbf{9.8} & \textbf{7.3} & \textbf{30.3} & \textbf{9.6} & 
    \textbf{4.9} & \textbf{17.7} & \textbf{66.6} \\
    \hline    
    \end{tabular}
    \end{threeparttable}
\end{table}

Table 3’s LAE-ST CTC model replaces the MoE layer in LAE-ST-MoE with a regular FFN. From the results, we can see that due to the introduction of the MoE module, the performance of ASR and ST is both improved obviously, which further confirms our motivation that introducing the MoE module
will make ASR and ST tasks more focused.

\subsubsection{Results of using ST decoder for ASR rescore}
\label{sssec:ab_use_st_for_asr}

\begin{table}[ht!] 
  \renewcommand\tabcolsep{4.7pt}
  \begin{threeparttable}
  \caption{Performance of using ST decoder rescore.
  } 
  \vspace{-1.5 em}
  
  \label{tab:res_st_rescore}
    \center

    \begin{tabular}{l | c c c | c | c} \hline
    \multicolumn{1}{c|}{\multirow{2}{*}{\textbf{Model}}} &
    \multicolumn{3}{c|}{\textbf{Code-Switch}} &
    \multicolumn{2}{c}{\textbf{Mono}}\\ \cline{2-6}
     & \textbf{ALL} &  \textbf{CN} & \textbf{EN} & \textbf{EN} & \textbf{CN} \\ 
     \hline  
    Vallina CTC  & 12.2 & 9.0 & 38.9 & 12.4 & 7.1 \\
    LAE CTC & 10.8 & 8.0 & 33.7 & 10.5 & 5.3 \\
    LAE-ST-MoE CTC & 9.8 & 7.3 & 30.3 & 9.6 & 4.9  \\
    ~ + En2Man ST Dec. res. & \textbf{9.7} & \textbf{7.1} & 31.2 & 10.2 & \textbf{4.8} \\
    ~ + Man2En ST Dec. res. & 10.4 & 8.1 & \textbf{29.1} & \textbf{9.3} & 5.6  \\
    \hline    
    \end{tabular}
    \end{threeparttable}
\end{table}

The En2Man ST decoder is comparable to the Mandarin ASR decoder for monolingual Mandarin input. Therefore, we can easily fuse it into monolingual Mandarin decoding through rescoring. As shown in Table~\ref{tab:res_st_rescore}, the En2Man ST decoder improves the LAE-ST-MoE CTC system in the mono CN speech. It achieves comparable results to the LAE-ST-MoE AED system (Table~\ref{tab:main_res}'s S6) in the monolingual Mandarin test. Especially on the Mandarin part of the CS test, the En2Man ST decoder rescoring performs better than the LAE-based AED system (Table~\ref{tab:main_res}'s S5), which maybe benefit from the Mandarin-English context representation and the decoder LM-related information. 
In addition, the same phenomenon also can be observed when applying the Man2En ST decoder rescoring.
These results show that the information learned by the ST decoder differs from that of the ASR decoder, improving the ASR performance. 
To a certain extent, the above results also prove the effectiveness of the LAE-ST-MoE.

\vspace{-0.5 em}

\subsubsection{Results of different $\rm \beta$ and $\mathbf {N_{Mono}}$ values in LAE-ST-MoE}
\label{sssec:ab_w_st_ps}
As mentioned in section~\ref{ssec:lae_st_moe_train_decode}, $\rm \beta$ is used to balance and regulate the ST effect. Therefore, in Table~\ref{tab:res_asr_beta}, we conduct
experiments with $\rm \beta$ values of 1.0, 0.8, 0.6, and 0.4, where we set $\mathbf{N_{Share}}$ = 9 and $\mathbf{N_{Mono}}$ = 1. From the results, it can be seen that the performance of CS is basically not affected, and the model has the best overall performance at 0.6.

\begin{table}[ht] 
  \renewcommand\tabcolsep{4.8pt}
  \begin{threeparttable}
  \caption{Results with different $\rm \beta$ 
  when $\mathbf{N_{Share}}$ = 9 and $\mathbf{N_{Mono}}$ = 1.
  } 
  \vspace{-1.0 em}
  \label{tab:res_asr_beta}
    \center
    \begin{tabular}{l c | c c c | c | c} \hline
    \multicolumn{1}{c}{\multirow{2}{*}{\textbf{Model}}} &
    \multicolumn{1}{c|}{\multirow{2}{*}{\textbf{$\mathbf {\beta}$}}} & 
    \multicolumn{3}{c|}{\textbf{Code-Switch}} &
    \multicolumn{2}{c}{\textbf{Mono}}\\ \cline{3-7}
     & & \textbf{ALL} &  \textbf{CN} & \textbf{EN} & \textbf{EN} & \textbf{CN} \\ \hline  
    Vallina CTC & - & 12.2 & 9.0 & 38.9 & 12.4 & 7.1 \\
    LAE-ST-MoE CTC & 1.0 & \textbf{9.8} & \textbf{7.3} & \textbf{30.3} & 9.7 & 5.0  \\
    LAE-ST-MoE CTC & 0.8 & \textbf{9.8} & \textbf{7.3} & \textbf{30.3} & 9.7 & 5.1  \\
    LAE-ST-MoE CTC & 0.6 & \textbf{9.8} & \textbf{7.3} & \textbf{30.3} & \textbf{9.6} & \textbf{4.9}  \\
    LAE-ST-MoE CTC & 0.4 & 9.9 & 7.4 & 30.5 & 9.7 &  5.0  \\
    \hline    
    \end{tabular}
    \end{threeparttable}
    \vspace{-0.5 em}
\end{table}

In addition, we set $\rm \beta$ to 0.6 and $\mathbf {N_{Share}}$ to 9. Then, the effectiveness of $\mathbf {N_{Mono}}$ is investigated in Table~\ref{tab:res_asr_N_mono}.
When $\mathbf {N_{Mono}}$ is 0, the ASR and ST share all encoder layers except FFN-MoE.
However, when $\mathbf {N_{Mono}}$=2, the LAE-ST-MoE encoder layer will reduce to 1. From Table~\ref{tab:res_asr_N_mono}, we can see the model achieves the best in $\mathbf {N_{Mono}}$=1, which suggests that the LAE-ST-MoE model needs more layers to perform ST, and it also needs to reserve some layers to learn the language-specific ASR representation. 

\begin{table}[ht] 
  \vspace{-1.5 em}
  \renewcommand\tabcolsep{3.5pt}
  \begin{threeparttable}
  \caption{Results with different $\mathbf{N_{Mono}}$ 
  when $\mathbf{N_{Share}}$ = 9 and $\rm \beta$ = 0.6. 
  } 
  \label{tab:res_asr_N_mono}
    \center
    \begin{tabular}{l c | c c c | c | c} \hline
    \multicolumn{1}{c}{\multirow{2}{*}{\textbf{Model}}} &
    \multicolumn{1}{c|}{\multirow{2}{*}{\textbf{$\mathbf {N_{Mono}}$}}} & 
    \multicolumn{3}{c|}{\textbf{Code-Switch}} &
    \multicolumn{2}{c}{\textbf{Mono}}\\ \cline{3-7}
     & & \textbf{ALL} &  \textbf{CN} & \textbf{EN} & \textbf{EN} & \textbf{CN} \\ \hline  
    Vallina CTC & - & 12.2 & 9.0 & 38.9 & 12.4 & 7.1 \\
    LAE-ST-MoE CTC & 0 & 10.1 & 7.5 & 31.6 & 9.9 & 5.1 \\
    LAE-ST-MoE CTC & 1 & \textbf{9.8} & \textbf{7.3} & \textbf{30.3} & \textbf{9.6} & \textbf{4.9}  \\
    LAE-ST-MoE CTC & 2 & 9.9 & 7.4 & 30.8 & 9.7 & 5.0 \\
    \hline    
    \end{tabular}
    \end{threeparttable}
    \vspace{-1.5 em}
\end{table}



\subsubsection{The results of ST auxiliary task in LAE-ST-MoE models}
\label{sssec:res_st}

\begin{table}[t] 
  \renewcommand\tabcolsep{2.0pt}
  \begin{threeparttable}
  \caption{ST results on the CS test 
  when $\mathbf{N_{Share}}$ = 9 and $\mathbf{N_{Mono}}$ = 1.} 
  \vspace{-1.5 em}
  \label{tab:res_st_res_alpha}
    \center
    \begin{tabular}{l c | c | c | c | c } \hline
    \multicolumn{1}{c}{\multirow{2}{*}{\textbf{Model}}} &
    \multicolumn{1}{c|}{\multirow{2}{*}{\textbf{$\rm \beta$}}} &
    \multicolumn{2}{c|}{\textbf{CS $\rightarrow$ EN}} &
    \multicolumn{2}{c}{\textbf{CS $\rightarrow$ CN}} \\ \cline{3-6}
    & & \textbf{BLEU} & \textbf{TER ($\downarrow$)} & \textbf{BLEU} & \textbf{TER ($\downarrow$)} \\
    \hline
    LAE-ST-MoE CTC  & 1.0 & \textbf{18.4} & \textbf{69.6} & \textbf{67.0} & \textbf{21.3}  \\
    LAE-ST-MoE CTC  & 0.8 & 18.1 & 70.0 & 66.8 & 21.5  \\
    LAE-ST-MoE CTC  & 0.6 & 17.7 & 70.3 & 66.6 & 21.6  \\
    LAE-ST-MoE CTC  & 0.4 & 17.3 & 70.6 & 66.2 & 21.9  \\
    \hline    
    \end{tabular}
    \end{threeparttable}
\end{table}

\begin{table}[t] 
  \renewcommand\tabcolsep{0.8pt}
  \begin{threeparttable}
  \caption{ST results on the CS test 
  when $\mathbf{N_{Share}}$ = 9 and $\rm \beta$ = 0.6.} 
  \vspace{-1.5 em}
  \label{tab:res_st_res_n_mono}
    \center
    \begin{tabular}{l c | c | c | c | c } \hline
    \multicolumn{1}{c}{\multirow{2}{*}{\textbf{Model}}} &
    \multicolumn{1}{c|}{\multirow{2}{*}{\textbf{$\mathbf {N_{Mono}}$}}} &
    \multicolumn{2}{c|}{\textbf{CS $\rightarrow$ EN}} &
    \multicolumn{2}{c}{\textbf{CS $\rightarrow$ CN}} \\ \cline{3-6}
    & & \textbf{BLEU} & \textbf{TER ($\downarrow$)} & \textbf{BLEU} & \textbf{TER ($\downarrow$)} \\
    \hline
    LAE-ST-MoE CTC  & 0 & \textbf{18.6} & \textbf{69.0} & \textbf{67.0} & \textbf{21.3}  \\
    LAE-ST-MoE CTC  & 1 & 17.7 & 70.3 & 66.6 & 21.6   \\
    LAE-ST-MoE CTC  & 2 & 16.3 & 72.6 & 65.5 & 22.4  \\
    \hline    
    \end{tabular}
    \end{threeparttable}
\end{table}

\begin{table}[ht!] 
  \renewcommand\tabcolsep{4.0pt}
  \begin{threeparttable}
  \caption{ST results on the monolingual test.
  }
  \vspace{-1.5 em}
  \label{tab:res_st_res_mono}
    \center
    \begin{tabular}{l | c | c | c | c } \hline
    \multicolumn{1}{c|}{\multirow{2}{*}{\textbf{Model}}} &
    \multicolumn{2}{c|}{\textbf{CN $\rightarrow$ EN}} &
    \multicolumn{2}{c}{\textbf{EN $\rightarrow$ CN}} \\ \cline{2-5}
    & \textbf{BLEU} & \textbf{TER ($\downarrow$)} & \textbf{BLEU} & \textbf{TER ($\downarrow$)} \\
    \hline
    LAE-ST-MoE CTC  & 33.9 & 44.8 & 31.5 & 59.1  \\
    \hline    
    \end{tabular}
    \end{threeparttable}
\vspace{-0.5 em}
\end{table}
 
We use  ModelScope's MT model to generate pseudo-labels for the test set. From Tables~\ref{tab:res_st_res_alpha} and \ref{tab:res_st_res_n_mono},
which show the BLEU score and translation error rate (TER) of our models, 
we can see that ST is less affected by $\rm \beta$ but more affected by $\mathbf {N_{Mono}}$. Furthermore, by combining Tables~\ref{tab:res_asr_beta}, \ref{tab:res_asr_N_mono}, \ref{tab:res_st_res_alpha}, and \ref{tab:res_st_res_n_mono}, we can observe that 
when the ST BLEU change, the ASR remain basically unchanged.
It may be because there is also some confusion between the information on ASR and ST. However, the helpful and confusing information needs to be balanced. 
Our experimental CS data is Mandarin-dominant, so we have more Mandarin-to-English ST training data than English-to-Mandarin, which results in better BLEU for Mandarin-to-English ST. 
Furthermore, we test the best ST model on monolingual data in Table~\ref{tab:res_st_res_mono}, and we can see that our model also has good BLEU. 
For CS data with limited English, the BLEU of 
CS speech to Mandarin text shows better than CS to English. 

\begin{figure}[t!]
\centering
 \centerline{\includegraphics[width=8.7cm]{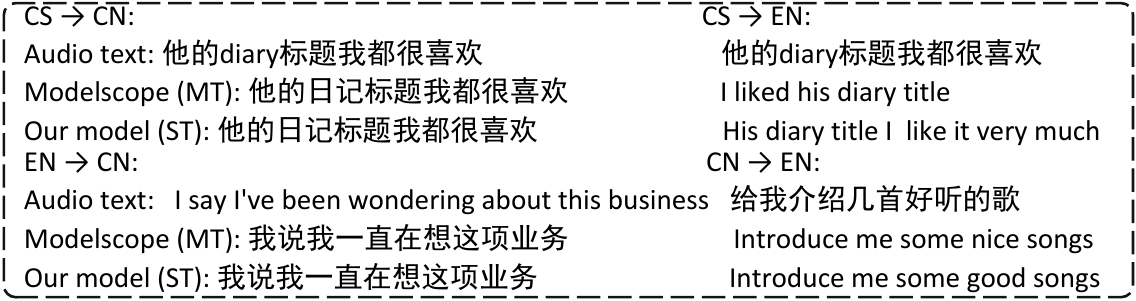}}
 \vspace{-1.0 em}
\caption{The examples translated by ModelScope and our model respectively.}
\label{fig:translate_case}
 \vspace{-1.5 em}
 
\end{figure}

Figure~\ref{fig:translate_case} provides examples of the translation performed by ModelScope’s MT model and our model's ST branch. Specifically, ModelScope's model translates text, whereas ours conducts the ST task. To a certain extent, these examples demonstrate that our model has learned good ST ability. 

  
    

\vspace{-0.5 em}

\section{Conclusions}
\label{sec:majhead}

In this paper, we propose an LAE-ST-MoE framework
that incorporates ST tasks into LAE and utilizes ST to learn the contextual information between different languages. 
The experimental results on the ASRU 2019 Mandarin-English CS challenge dataset demonstrate that, 
compared to the LAE-based CTC and AED system, the proposed LAE-ST-MoE model achieves about 6\%-9\% relative error rate reduction. 
Extensive investigations into the w/ or w/o MoE module, comparison with the literature results, and ablation on different $\rm \beta$ and $\mathbf {N_{Mono}}$ values have also been carried out and confirm the effectiveness of the 
LAE-ST-MoE.
Moreover, the well-trained LAE-ST-MoE model can perform ST tasks from CS speech to Mandarin or English, and the structure is easy to extend to one-to-many ST tasks. 
In the future, we will further explore the LAE-ST-MoE to multilingual ASR and one-to-many ST. 

\vfill\pagebreak 

\bibliographystyle{IEEEbib}
\bibliography{strings,refs}

\end{document}